# Performance of Centroiding Algorithms at Low Light Level Conditions in Adaptive Optics

Akondi Vyas, M B Roopashree, B R Prasad
Laser Lab, CREST,
Indian Institute of Astrophysics
Bangalore, India

Akondi Vyas
Physics Department
Indian Institute of Science
Bangalore, India

*Abstract*—The performance metrics of different centroiding algorithms at low light level conditions were optimized in the case of a Shack Hartmann Sensor (SHS) for efficient performance of the adaptive optics system. For short exposures and low photon flux, the Hartmann spot does not have a Gaussian shape due to the photon noise which follows poissonian statistics. The centroiding estimation error was calculated at different photon levels in the case of changing spot size and shift in the spot using Monte Carlo simulations. This analysis also proves to be helpful in optimizing the SHS specifications at low light levels.

## I. Introduction

Adaptive Optics (AO) is a real time wave-front correcting technology that compensates for the continuous fluctuations due to turbulent medium [1]. The three major components in a simple AO system are wave-front sensor, wave-front corrector and the control algorithm. A wave-front sensor helps in determining the shape of the incoming distorted wave-front. The control algorithm reads the wave-front sensor output and calculates the conjugate wave-front that is applied on the wave-front correcting system.

Most widely used wave-front sensor in AO is the Shack Hartmann Sensor (SHS). The basic working of the SHS is depicted in Fig. 1.

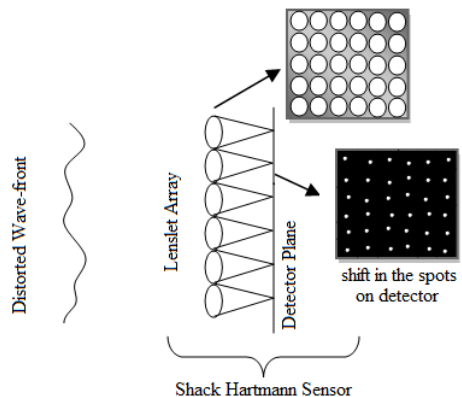

Fig. 1. Working of a Shack Hartmann Sensor

It consists of a two dimensional array of lenslets and a detector placed at the focal plane of the lenslet array. Each lenslet is called the subaperture of the array. When light of sufficient intensity is incident on the SHS, it forms Gaussian like spots on the detector referred to as spots in this paper. The positions of the spots change in accordance with the local wave-front gradients. Centroiding is performed on areas corresponding to individual lenslets on the detector to figure out the position of the spots. Comparing with a standard reference, the wave-front is reconstructed through wave-front reconstruction algorithms.

In the case of astronomical adaptive optics, the problem of low light level conditions is a frequently encountered problem [2]. Since the object under observation is usually faint, a guide star is used for wave-front sensing purposes. The guide star which is a natural star that is close to the object under observation and sufficiently bright is called the Natural Guide Star (NGS). A guide star that is generated artificially using high power lasers is called the Laser Guide Star (LGS). In the absence of LGS system and a sufficiently bright NGS, we are forced to choose a star which is not sufficiently bright and which is close to the object under observation. If the star light is low, the exposure time needs to be large and which will add to the servo lag errors in AO. Hence the exposure time cannot be increased indiscriminately large. This is why it becomes crucial to study the centroid detection problem at low light levels [3]. Also the centroiding estimation error which is defined later in the paper needs to be significantly low in astronomical cases since the local phase fluctuations have very low amplitudes so that they give rise to subpixel shift in the spots.

Multiple centroiding techniques are extensively discussed in the literature. Center of Gravity (CoG), Weighted CoG (WCoG), Iteratively WCoG (IWCoG), matched filtering and correlation based centroiding techniques are out of the well studied algorithms [4-6].

CoG is the simplest of all the techniques with low computational effort. WCoG has an advantage over CoG since it takes into consideration the fact that the spots attain a Gaussian structure. Hence using a Gaussian weighting function, it is more probable that the estimated centroid matches with the actual one. In IWCoG, the weighting function's width, and centre are corrected iteratively and the centroid is estimated. This technique takes more computational time as the number of iterations is increased. WCoG can also be performed by using intensity function itself as the weighting function, in which case, the technique is called Intensity Weighted Centroiding (IWC).

The parameters internal to SHS that might affect the centroiding accuracy include the detector size, SHS pitch (distance between two lenslets), and spot size. The factors external to SHS affecting the centroid estimation are the photon flux (number of incident photons per unit area per unit time), exposure time, shift in the spot from the centre of the subaperture and centroiding technique used.

The detector size and the SHS pitch decide the number of pixels that correspond to a single subaperture on the CCD (most generally used detectors with SHS). Reducing the SHS pitch reduces the space allotted to individual subapertures on the CCD. Spot size reduction is beneficial in the case of low photon level conditions since tight focusing helps intensity build up by photon addition in the region around the centroid.

The number of photons per second, $N_{photons}$ falling on a single subaperture of diameter, 'd' is given by,

$$N_{photons} = F_{Star}\, d^2 \quad (1)$$

where $F_{Star}$ is the photon flux of the star in photons/m$^2$/sec. A magnitude 10 star can be used as a NGS since a few hundred photons can be supplied by the star to each subaperture in the case of 1m telescope with 1000 subapertures even after including optical transmission as well as detector inefficiencies. Using higher magnitude stars reduces the available photons and hence makes detection of the centroid difficult in a SHS.

Exposure time has a direct relation to the number of photons collected in a single shot. Although atmosphere is dynamic, it maintains correlation within a few milliseconds and the time over which it completely loses information of its past is called the decorrelation time [7]. In AO, wave-front correction has to be done within this decorrelation time scale. Hence the exposure time has to be much smaller than the decorrelation time of turbulence. Exposure time may vary from 1millisecond to tens of milliseconds. It is generally fixed by looking at the decorrelation time and the photon flux.

In this paper we studied the performance of CoG, WCoG, IWCoG, IWC techniques in the presence of low light level conditions. The effect on the centroid estimation at low light was examined at different spot sizes, shifts in the spot and CCD sizes. The second section defines the different centroiding techniques and presents a comparative analysis of their performance in the presence of a uniform background noise. The third section describes the method adopted to simulate spots influenced by the existence of photon noise. This is done by controlling the number of photons falling on the subaperture. In the fourth section, computational results have been presented. In the last section the conclusions are presented.

II. CENTROIDING ALGORITHMS

In the absence of noise, CoG gives the exact centroid positions and will not have any noise related errors. But, in the presence of noise, each of the centroiding techniques has advantages and disadvantages. This fact is clearly evident from the performance analysis of the centroiding algorithms in the presence of a uniform background noise which was studied earlier [8]. Hence it becomes important to study the performance of these methods in the presence of Poisson noise too.

*A. Center of Gravity (CoG)*

CoG technique identifies the geometric center of the object shape through global averaging. Let us consider a two dimensional image, I(x, y) with M×M pixels and discrete intensity values at coordinates $X_{ij} = (x_i, y_j)$, where $i, j$ take values 1,2,....M. Then the centroid, $(x_c, y_c)$ as defined by CoG is given by,

$$(x_c, y_c) = \frac{\sum_{ij} X_{ij} I_{ij}}{\sum_{ij} I_{ij}} \quad (2)$$

This is the simplest of the centroiding techniques and is easy to compute. The difficulty in computation increases with increasing image size and generally the subaperture size is kept at a decently low value. The application of CoG in the presence of a uniform background is not useful.

*B. Weighted Center of Gravity (WCoG)*

A SHS spot ideally looks like an airy pattern and can be easily approximated to a two dimensional Gaussian function. Taking the advantage of the shape of the spot, we weigh the centroid estimation formula with the weighting function.

$$(x_c, y_c) = \frac{\sum_{ij} X_{ij} I_{ij} W_{ij}}{\sum_{ij} I_{ij}} \quad (3)$$

where, W(x,y) is the Gaussian weighting function with a spread given by σ.

$$W(x, y) = \frac{1}{2\pi\sigma^2} \exp\left\{-\frac{(x - x_c)^2}{2\sigma^2} - \frac{(y - y_c)^2}{2\sigma^2}\right\} \quad (4)$$

This method has an advantage over CoG when the Gaussian spot shape is maintained and if the spot does not shift by a large amount from the image center. It is disadvantageous and sometimes even useless in the presence of strong noise background or even in conditions of barely few photons.

*C. Iteratively Weighted Center of Gravity (IWCoG)*

Since the weighting function in Eq. 4. depends on the position of the centroid and the spot size which is not in our control, an iterative process is followed. As an initial guess, the image center is taken as the centroid estimate and one quarter of the image length is taken as the spot width. In the iterations, spread in the spot and the position of the centroid are progressively corrected after every iteration. For further detailed understanding refer to [5].

IWCoG gives an accurate centroid estimate compared to WCoG and CoG even at large shift in the spots. It is computationally more intense compared to any other centroiding algorithm. Increasing the number of iterations may decrease the centroiding error, but increases the computational time. Hence it is important to optimize for the number of iterations to be used for best performance.

### D. Intensity Weighted Centroiding (IWC)

IWC is similar to WCoG with the centroid given by Eq. 3, with a difference that the weighting function is intensity dependent

$$W(x, y) = I^p(x, y) \quad (5)$$

where '$p$' can take any real positive number greater than one. Mathematically, intensity weighted centroiding with $p=1$ is the optimum technique in the presence of noise.

### III. SIMULATION OF SPOTS WITH POISSON NOISE

Unlike readout noise, background noise and amplifier noise, photon noise is a signal dependent noise and hence cannot be avoided. If we monitor an individual pixel of the CCD, there will be a finite probability for a photon to arrive within a given time interval. This random process can be described mathematically by a Poisson process.

There are many methods to simulate Poisson noise on images in the literature. We adopted a very simple method to generate photon depleted spots. A SHS spot formed out of sufficient number of photons forms a Gaussian shape. As a first step, we simulate a Gaussian spot intensity G (x, y) of known spread and center. As a next step, two random vectors X, Y of lengths equal to the number of incident photons are generated. These vectors are realizations from a uniform random sample. In our application, these vectors are the position coordinates where the photons are incident. As a third step, the intensity at these coordinates is given by, I(X, Y) = G(X, Y). In this manner, the position coordinates of the photon hits are determined and the intensity corresponding to the Gaussian spot is addressed to the new image.

An illustration of the simulated Poisson noise with different photon levels is shown in Fig. 2.

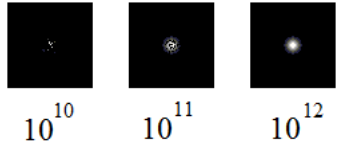

Fig. 2. SHS spot at different photon flux (photons/m$^2$/sec), worked out for 1m telescope, 5ms exposure, 1000 subapertures

The Centroid Estimation Error (CEE) depends on how far the estimated centroid is from the actual centroid. It is defined by using the following formula,

$$CEE = \sqrt{(x_c^* - x_c)^2 + (y_c^* - y_c)^2} \quad (6)$$

where, ($x_c^*$, $y_c^*$) is the estimated centroid and ($x_c$, $y_c$) is the actual centroid. It represents the distance between the actual centroid and the position of the estimated centroid. CEE hence has the units of pixels. In all of the simulations in this paper, the exposure time was fixed at 5 milliseconds. A 1m class telescope was chosen with 1000 subapertures. Hence if the photon flux is $10^{10}$ photons/m$^2$/sec, then excluding the effects of readout noise, quantization noise and the optical transmission inabilities, $5\times10^4$ photons must be received at each subaperture. Including these losses, this number drops by nearly 2 orders of magnitude.

### IV. COMPUTATIONAL RESULTS

The spot images were simulated with known photon flux and the performance of different centroiding techniques was tested at different parameters like spot size, shift in the spot and CCD size using Monte Carlo simulations.

### A. Effect of changing spot size

As the spot size increases, the photons have to be distributed in a larger area and hence individual photons will have larger spatial choice. This will lead to larger errors in the case of low photon count. In the case of smaller, focused spots, the photons tend to concentrate over a much smaller area and hence can easily form a Gaussian even with less number of photons.

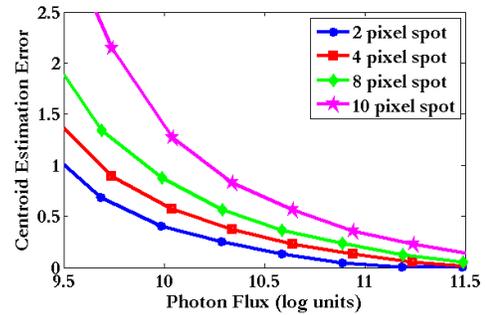

Fig. 3. Effect of changing spot size on the CEE at different photon flux.

The effect of changing spot size is shown in Fig. 3. It is clear that at low photon flux, the CEE is worse and gets better for higher photon flux. Although the centroiding technique used here is IWCoG with 8 iterations, a similar behavior was observed in the case of all the methods.

### B. Effect of shift in the spot

Depending on the local gradient of the wave-front, the spot shifts from the center of the subaperture. The presence of Poisson noise over the shifted spots does not add any special significance to the centroiding accuracy of different techniques. That can be seen in Fig. 4. where for different magnitudes of shift in the spot, the performance of CoG and IWC remains unaltered.

WCoG method fails to accurately find out the centroid in the case of shift in the spot since the initial guess for spot center and the spread is not the right one. IWCoG with more number of iterations takes it closer to the actual centroid as shown in Fig. 5.

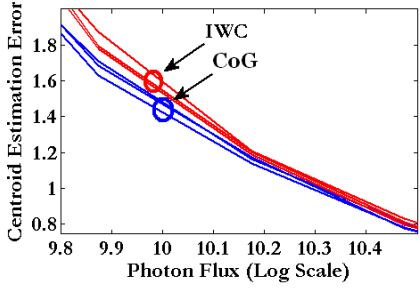

Fig. 4. IWC and CoG methods remain unchanged with different magnitudes of shift in the spots

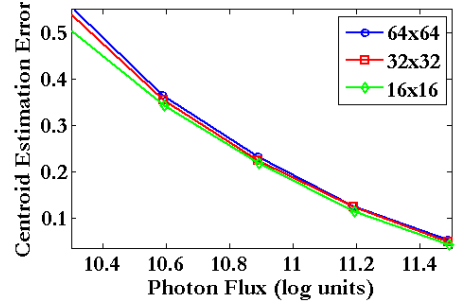

Fig. 7. Effect of changing CCD size on centroiding using IWCoG

## V. DISCUSSION AND CONCLUSIONS

The performance metrics were defined that govern the centroiding accuracy in a SHS at low light level conditions in astronomical AO. For efficient performance of the adaptive optics system, it is very important to optimize the computational, experimental as well as device parameters like the centroiding algorithm used, the shift in the SHS spots, the exposure time, photon flux, the spot size and the CCD size.

It was seen that maintaining a small spot size will help us in keeping the centroiding error to a low value. Since WCoG gives large errors with shift in the spots, it must be avoided and replaced by IWCoG technique. Better optimization of the number of iterations is necessary for best utilization of time and efficiency. CoG method and IWC method perform best at low light levels when comapared to IWCoG. IWCoG performs better at high intensity levels. Intensity weighted works best with weighting power, p=1. Reducing the CCD size reduces the CEE. The analysis proves helpful in optimizing the SHS specifications at low light levels.

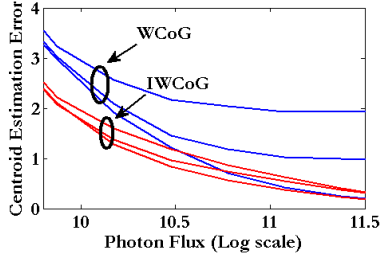

Fig. 5. Performance of WCoG and IWCoG with shift in the spot

In the case of WCoG, for a shift of 2 pixels, the CEE is above 2 even at higher photon levels. For 1 pixels shift it still stays above CEE=1. Above a photon flux $10^{11}$ /m$^2$/s, in the case of IWCoG (a maximum of 8 iterations), CEE drops below 1.

### C. Comparison of centroiding algorithms

Although the CEE does not differ by a very large amount in each of the techniques, but CoG, IWC methods dominate at low flux conditions.

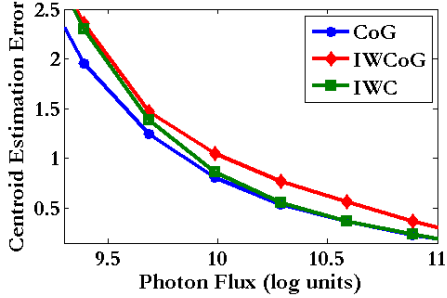

Fig. 6. Comparison of centroiding algorithms

### D. Effect of changing CCD size

Our analysis shows that changing the CCD size does not have a significant effect on the centroiding accuracy in the presence of Poisson noise. For spot size of 4 pixels and using IWCoG with 8 iterations, the CEE was calculated at different photon levels. This was done for different subaperture size or CCD size. The CCD size was varied - 64×64, 32×32 and 16×16 and the CEE is plotted as a function of photon flux in Fig. 7.